\documentclass{WileyMSP-template}
\usepackage{xcolor}
\begin{document}

\pagestyle{fancy}
%\rhead{\includegraphics[width=2.5cm]{vch-logo.png}}

\justify

\title{Patterning of superconducting two-dimensional electron gases based on AlO$_x$/KTaO$_3$(111) interfaces}
\maketitle

% Author: Please give full first and last names for authors and include * after the name of all corresponding authors
\noindent
\author{Hugo Witt}
\author{Srijani Mallik}
\author{Luis Moreno Vicente-Arche}
\author{Gerbold Ménard}
\author{Guilhem Saïz}
\author{Daniela Stornaiuolo}
\author{Maria D'Antuono}
%\author{Marco Salluzzo}
\author{Isabella Boventer}
\author{Nicolas Bergeal}
\author{Manuel Bibes*}

% Dedication
% \dedication{Optional dedication here. If no dedication is required, please leave blank}
\dedication{}

% Affiliations: Please provide academic titles (Prof. or Dr.) for all authors where applicable, and include an institutional email address for all corresponding authors
\begin{affiliations}
H. Witt, Dr. S. Mallik, Dr. L. M. Vicente-Arche, Dr. I. Boventer, Dr. M. Bibes\\
Unité Mixte de Physique, CNRS, Thales, Université Paris-Saclay, 1 Avenue Augustin Fresnel, 91767 Palaiseau, France\\
*Email Address: manuel.bibes@cnrs-thales.fr\\
\hfill \break
H. Witt, Dr. G. Ménard, Dr G. Saïz, Dr. N. Bergeal\\
Laboratoire de Physique et d’Etude des Matériaux, ESPCI Paris, Université PSL, CNRS, Sorbonne Université, Paris, France\\
\hfill \break
Dr. D. Stornaiuolo, Dr. M. D'Antuono\\
CNR-SPIN, Napoli, Italy\\
Dipartimento di Fisica, Università di Napoli “Federico II”, Napoli, Italy\\
\end{affiliations}

% Keywords: Please provide a minimum of three and a maximum of seven keywords, separated by commas

% Abstract should be written in the present tense and impersonal style (i.e., avoid we), and be at most 200 words long

\begin{abstract}
The versatility of properties displayed by two-dimensional electron gases (2DEGs) at oxide interfaces has fostered intense research in hope of achieving exotic electromagnetic effects in confined systems. Of particular interest is the recently discovered superconducting state appearing in (111)-oriented KTaO$_3$ interfaces, with a critical temperature \textit{T$_c$}$ \approx 2$~K, almost ten times higher than that of SrTiO$_3$-based 2DEGs. Just as in SrTiO$_3$-based 2DEGs, fabricating devices in this new system is a technical challenge due to the fragility of the 2DEG and the propensity of bulk KTaO$_3$ to become conducting outside the devices upon adventitious oxygen vacancy doping. Here, we present three different techniques for patterning Hall bars in AlO$_x$/KTaO$_3$~(111) heterostructures. The devices show superconducting transitions ranging from 1.3~K to 1.78~K, with limited degradation from the unpatterned thin film, and enable an efficient tuning of the carrier density by electric field effect. The array of techniques allows for the definition of channels with a large range of dimensions for the design of various kinds of devices to explore the properties of this system down to the nanoscale.
\end{abstract}

% Text: Please use section headings and subheadings as specified below. For communications, all section headings apart from Experimental Section should be removed
% Please make the first reference to a display item bold: \textbf{Figure 1}
% Do not abbreviate Figure, Equation, etc.; display items are always singular, i.e., Figure 1 and 2.
% Equations are always singular, i.e., Equation 1 and 2, and should be inserted using the {equation} environment, not as graphics
% Please do not use footnotes in the text, additional information can be added to the Reference list.

\vspace{2em}

%\section{Introduction}

The study of heterostructures based on KTaO$_3$ gained significant interest with the discovery of superconductivity in two-dimensional electron gases (2DEG) at interfaces between KTaO$_3$ and various overlayers \cite{liu_two-dimensional_2021, chen_electric_2021, mallik_superfluid_2022}. Taking inspiration from the research on SrTiO$_3$ interfaces \cite{yun-yi_pai_physics_2018}, KTaO$_3$-based 2DEGs could serve as the main building block for devices with applications in spintronics \cite{trier_oxide_2022}, orbitronics \cite{johansson_spin_2021,vicente-arche_spincharge_2021, varotto_direct_2022}, and topological quantum computing \cite{barthelemy_quasi-two-dimensional_2021}. Indeed, with a superconducting critical temperature (\textit{T$_c$}) and a Rashba splitting of an order of magnitude higher in KTaO$_3$~(111) \cite{liu_two-dimensional_2021,varotto_direct_2022} compared to SrTiO$_3$ \cite{reyren_superconducting_2007,caviglia_tunable_2010,vaz_determining_2020}, the functional perspectives of this material look promising. However, the discovery and control of the electronic properties of KTaO$_3$-based interfaces, and all the more so of KTaO$_3$~(111), are hampered by difficulties in device fabrication. Indeed, on the material side, the KTaO$_3$ surface is sensitive, with volatile K cations and O anions that can be lost by annealing, etching or polishing. Regarding processing, the use of classic ultraviolet (UV) lithography polymer resist during growth is excluded by high temperature steps that are integral parts to the 2DEG formation process \cite{krantz_emergent_2022}. The tunability of transport properties is also limited by the lower dielectric constant of KTaO$_3$ than that of SrTiO$_3$ \cite{wemple_transport_1965}. 

Hitherto, the majority of the studies of KTaO$_3$ that show superconductivity have been performed on unpatterned thin film samples \cite{liu_two-dimensional_2021, chen_electric_2021, mallik_superfluid_2022, al-tawhid_coexistence_2021, liu_tunable_2022, sun_effects_2022}. Nonetheless, a variety of techniques have been proposed to design Hall bars: wet HCl etching on EuO/KTaO$_3$ (111) \cite{ma_superconductor-metal_2020, qiao_gate_2021}, etching on YAlO$_3$/KTaO$_3$~(111) \cite{zhang_spontaneous_2021}, scratching on Al$_2$O$_3$/KTaO$_3$~(111) \cite{ojha_anomalous_2022}, conducting AFM charging \cite{yu_nanoscale_2022} and ultra-low voltage electron beam lithography \cite{yu_nanoscale_2022} as well as using a thick amorphous Al$_2$O$_3$ hard mask deposited by pulsed laser deposition \cite{chen_electric_2021} on LaAlO$_3$/KTaO$_3$~(111) and LaAlO$_3$/KTaO$_3$~(110), or resorting to ionic liquid gating on KTaO$_3$~(111) \cite{ren_two-dimensional_2022}. 

In this study, we report the fabrication of Hall bars in superconducting KTaO$_3$~(111) 2DEGs patterned employing three methods: (i) stainless steel shadow masking, (ii) Ar$^+$ beam milling at cryogenic temperature, and (iii) using an insulating Al$_2$O$_3$ hard mask deposited by sputtering. Combined with a simple Al metal deposition method to create the 2DEG \cite{mallik_superfluid_2022}, these patterning recipes may be easily implemented using classic lithography capabilities. We demonstrate a fine tuning control of the carrier density by electric field effect, modulated by the width of the channel. 
\\\
%\section{Methods}

For all samples the 2DEG was formed by growing a 1.8 nm Al film at 500°C by DC magnetron sputtering system (PLASSYS MP450S), as described in Ref. \cite{mallik_superfluid_2022}, with substrates supplied by MTI corporation. The characterisation was conducted by optical and atomic force microscopy, preliminary transport measurements were done in a Quantum Design Dynacool system and low temperature transport measurements in a dilution refrigerator from 30~mK to 300~K.
\\\

%\section{Results}

%\subsection{Shadow mask}

\begin{figure}[h!]
    \label{fig:shadow mask}
    \includegraphics[width=\linewidth]{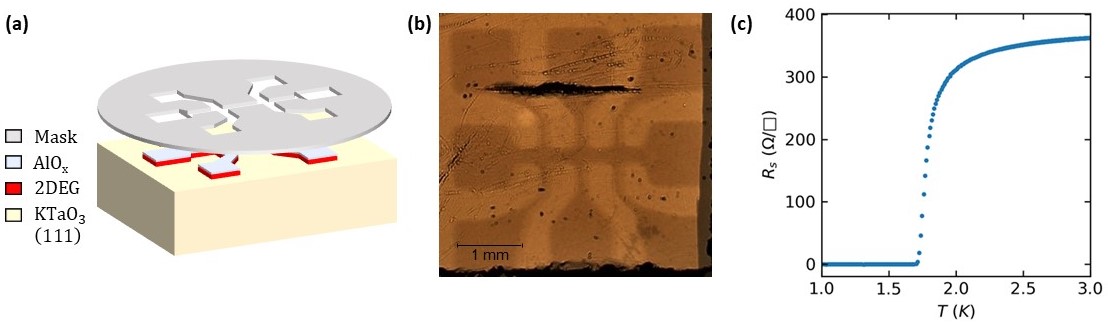}
    \caption{\textbf{Shadow mask method} (a) Sketch of the fabrication principle. The stainless steel mask is brought as close as possible to the substrate to minimize the shadow effect. (b) Optical microscopic view of the 250~µm Hall bar. The horizontal black line is a scratch on the substrate back side used to keep track of the crystallographic orientation of the substrate. (c) Low-temperature dependency of the longitudinal resistivity showing the superconducting transition at 1.78~K.}
\end{figure}

The first method we employed to pattern a Hall bar on KTaO$_3$~(111) substrates was shadow masking with a stainless steel mask. The advantage of this technique is the possibility to minimize the chemical impact on the surface of the as-received substrate. However, the dimensions of the features are limited by the resolution of the steel cutting technique. 

We used a 150~µm thick laser cut stainless steel mask in a 4~mm~x~4~mm design of a 250~µm wide and 2~mm long Hall bar, mounted \textit{ex situ} on the substrate plate without prior treatment to the substrate. A thin layer of Al was then sputtered on the KTaO$_3$ substrate, as described in Figure 1~(a) and Ref.~\cite{mallik_superfluid_2022}. The shadow effect is negligible compared with the dimensions of the device and channels are well defined (Figure 1~(b)). Atomic force microscopy (not shown) revealed a smooth surface on the Hall bar, comparable to thin film samples. X-ray photoelectron spectroscopy showed no contamination by the mask.

Figure 1~(c) displays the low-temperature dependence of the sheet resistance. We observe a metallic behavior characteristic of a 2DEG,  with a superconducting transition at \textit{T$_c$}$ \approx 1.78$~K, replicating the behaviour of an unpatterned thin film \cite{mallik_superfluid_2022}.
\\\
%\subsection{\label{sec:level1}Low temperature etching}

\begin{figure}[h]
    \label{fig:lowT etching}
    \includegraphics[width=\linewidth]{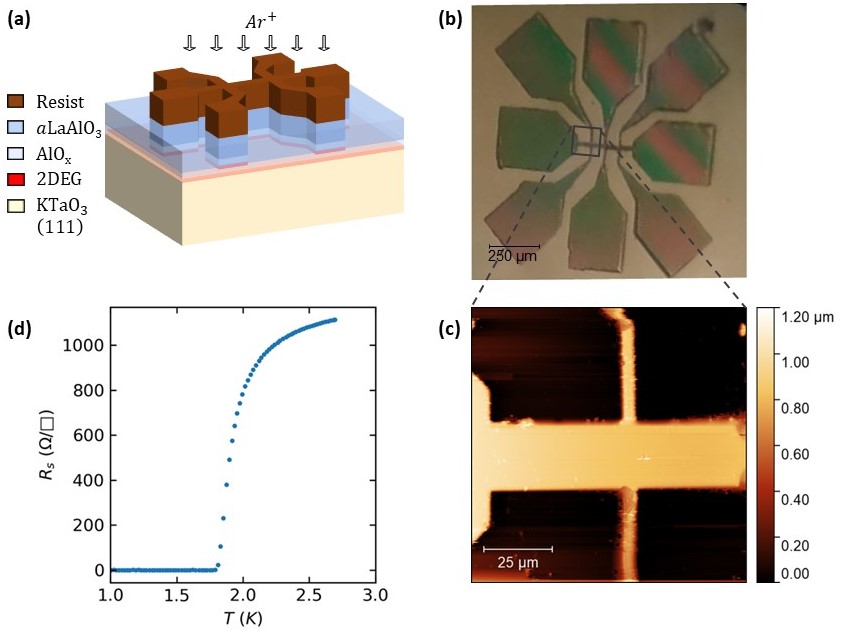}
    \caption{\textbf{Ar$^+$ beam milling method}(a) Sketch of the sample in the final step of fabrication. The transparent layers, not protected by the resist pattern, are etched during the low-temperature milling process, down to a few nanometers into the bare substrate. (b) Optical and (c) atomic force microscopic view of the 25~µm Hall bar. (d) Low-temperature dependance of the longitudinal resistance showing the superconducting transition at 1.8~K.}
\end{figure}

The second method yielding a superconducting 2DEG channel at the AlO$_x$/KTaO$_3$~(111) interface is liquid N$_2$-cooled argon ion beam etching. Ion beam etching is seldom used to pattern 2DEG-hosting interfaces due to the formation of oxygen vacancies at the substrate’s surface upon ion irradiation \cite{kan_blue-light_2005, herranz_vacancy_2010}. The vacancies dope the substrate with electrons, shunting the 2DEG in transport. KTaO$_3$ is anologous to SrTiO$_3$ in this regard \cite{ojha_oxygen_2021}. To avoid this effect on the LaAlO$_3$/SrTiO$_3$ system, a low energy ion beam can be used to suppress conductivity by turning the LaAlO$_3$ amorphous without physically etching it, avoiding SrTiO$_3$ irradiation \cite{paolo_aurino_nano-patterning_2013}. Another approach consists in reducing the kinetics of the oxygen vacancy formation and diffusion out of the sample by cooling it with liquid nitrogen during the ion beam etching \cite{dantuono_nanopatterning_2022}. The sides of the devices thus fabricated are free, and could couple in-plane to subsequently grown complex materials such as conventional superconductors. Using this technique, a pattern resolution of 160~nm could be reached \cite{dantuono_nanopatterning_2022}.

Here, a thin film of Al was sputtered on a (111)-oriented KTaO$_3$ and capped by a protective layer of amorphous LaAlO$_3$ subsequently grown by pulsed laser deposition. We then patterned the sample by UV lithography to form 25 µm wide Hall bars. The resist used was SPR~700~1.0, developed with MF-319, creating a 1.1 µm thick resist mask on top of the 2DEG. The system was then etched by a low-intensity Ar ion beam with a current of 5~mA and beam voltage of 315~V, resulting in an etching rate of 1~nm/s. During etching, the sample was glued to a cold finger kept at 77~K. Testing the resistance of the etched surface immediately after etching confirmed the absence of conductivity outside of the devices. 

Optical and AFM microscopy confirms the structural integrity of the channels thus defined down to 5~µm for the contact channels on the measured samples (Figure 2~(b) and ~(c)), only limited by the precision of the UV lithography.

The device has a characteristic behavior of a metallic 2DEG where the resistance decreases with the temperature. At low temperature, the resistivity measurements show a superconducting transition at \textit{T$_c$}$ \approx 1.8~$K (Figure 2~(d)), identical to the \textit{T$_c$} obtained for a thin film \cite{mallik_superfluid_2022}. 
\\\
%\subsection{Hard mask method}

\begin{figure}[h]
    \label{fig:Mannhart}
    \includegraphics[width=\linewidth]{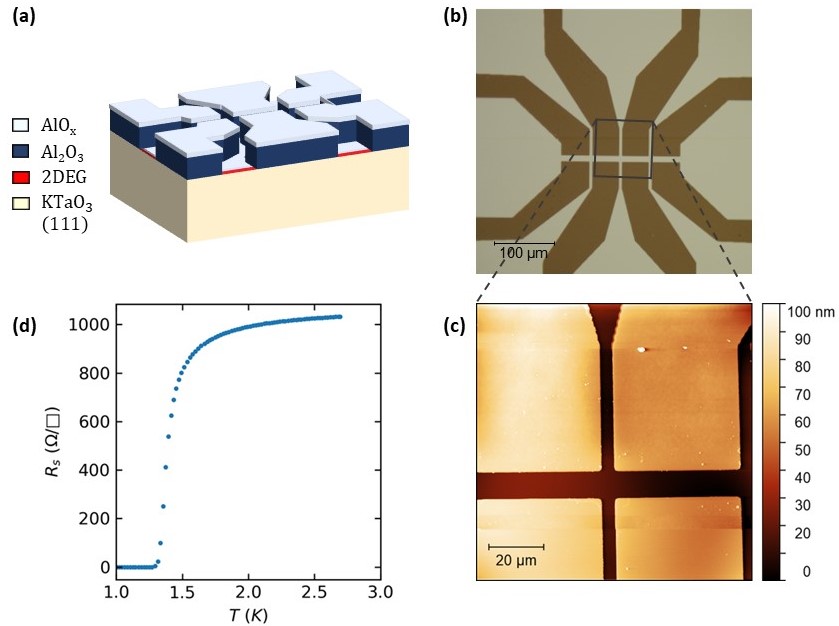}
    \caption{\textbf{Hard mask method} (a) Sketch of the sample in the final state of fabrication. The AlO$_x$ layer topping the Al$_2$O$_3$ hard mask is the layer of metallic Al, deposited together with what forms the 2DEG at the interface, that oxidizes in an insulating layer in contact with the mask. (b) Optical and (c) atomic force microscopic view of the 10~µm Hall bar. (d) Low-temperature dependency of the longitudinal resistance showing the superconducting transition at 1.3~K.}
\end{figure}

The third approach is based on the method developed by Schneider \emph{et al.} \cite{schneider_microlithography_2006} and consists in restricting the surface available for 2DEG formation by first patterning a hard mask on the sample using UV lithography. The desired design for the device is patterned in resist, directly on the substrate. A layer of insulating material is then grown on the sample at room temperature, such that, after lift-off, it forms a template for the 2DEG to be grown onto (Figure 3~(a)). This hard mask then prevents the formation of a conducting layer outside of the pattern.
Versatile insulators can be used as hard mask such as MgO, amorphous LaAlO$_3$ or Al$_2$O$_3$, using different deposition techniques including magnetron sputtering or pulsed laser deposition, in conditions ensuring that no conducting layer forms in the substrate underneath. This technique allows the use of simple UV lithography and lift-off processes despite the high temperature annealing and growth. Another advantage is the possibility to see the devices with the naked eye thanks to the contrast between the thick opaque hard mask layer and the thin and transparent conducting regions. However, the contact between the KTaO$_3$ surface and the resist prior to the growth of the 2DEG may degrade the quality of the substrate surface, despite  cleaning and annealing steps following the lift-off, as opposed to a deposition on a fresh substrate.

For this third method, we patterned the KTaO$_3$ sample with resist using the negative image of the mask that was used for the ion beam etching method. A 50~nm thick layer of amorphous Al$_2$O$_3$ was sputtered with 30~W power in a $5 \times 10^{-4}$~mbar Ar atmosphere. After lift-off, Al was sputtered on the sample with the same recipe as the other patterning techniques. The Al in contact with the substrate, in the areas not covered by the hard mask, forms the 2DEG, while the film deposited on the hard mask gets oxidized by the oxide insulator itself and by air \textit{ex situ}. We verified that there was no conducting metallic Al bridge between devices over the mask, that would induce leakage.

Microscopic imaging (Figure 3~(b)) shows a large contrast between the amorphous Al$_2$O$_3$ hard mask and the transparent 2DEG device. The AFM scan (Figure 3~(c)) of the 10~µm wide channel reveals a smooth surface for both the hard mask and the AlO$_x$/KTaO$_3$ interface and confirms the good definition of the 5~µm wide contact channels.
We performed temperature dependent resistance measurement to assess superconductivity, confirming the two-dimensional nature of the conducting layer with the characteristic metallic behaviour, and the transition at  \textit{T$_c$}$ \approx 1.3$~K (Figure 3~(d)). 
\\\ 

%\subsection{Gating experiments}

\begin{figure*}[h]
    \label{fig:density}
    \includegraphics[width=\linewidth]{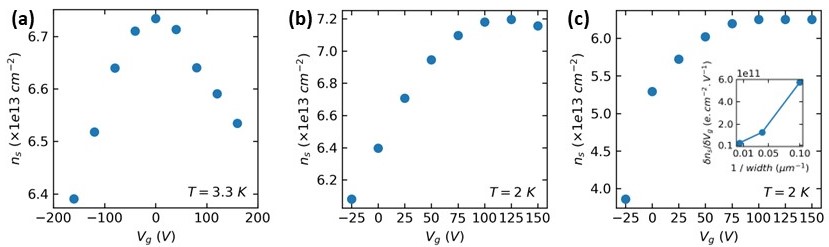} %\includegraphics[width=\linewidth]{density_modulation.jpg}
    \caption{\textbf{Carrier density tuning} Field effect carrier density modulation from Hall measurements with a back gate voltage, assuming a single type of charge carriers, of the three samples presented : (a) 250 µm wide Hall bar deposited by shadow mask, (b) 25 µm wide Hall bar etched at low temperature, (c) 10 µm wide Hall bar in an amorphous Al$_2$O$_3$ hard mask. (inset) Dependency of the maximum gradient of carrier density modulation with the inverse of the width, for the three devices studied.}
\end{figure*}

We conducted Hall measurements on all three samples while applying a back gate voltage. The Hall effect was linear for all samples at all gate voltages. We extracted the carrier density assuming a single type of carriers \cite{bruno_band_2019}. All samples had carrier densities of about $6 \times 10^{13}$ cm$^2$ in the virgin state at 2~K, before any gate application. The mobilities were in the $130-250$ cm$^2$~V$^{-1}$~s$^{-1}$ range, which is in the high range of reported values \cite{liu_two-dimensional_2021, zhang_spontaneous_2021, ojha_anomalous_2022, liu_tunable_2022, liu_superconductivity_2022, sun_effects_2022}. Figure 4 shows the gate dependency of the carrier density for the three samples. The amplitude of the carrier density modulation is inversely proportional to the channel width, varying over the explored gate voltage range by $0.34 \times 10^{13}$~cm$^{-2}$  in a 250~µm wide channel made by the shadow mask method, $1.11\times 10^{13}$ cm$^{-2}$ i.e. one order of magnitude higher in a 25~µm wide channel made by low-temperature etching and $3.43 \times 10^{13}$ cm$^{-2}$ in a 10~µm wide channel in an insulating hard mask.
The field effect efficiency, defined as the maximum of the gradient of the carrier density dependency on gate voltage, increases proportionally with the inverse of the channel width as shown in the inset in Figure 4~(c), as charges accumulate on a smaller area, leading to a large electric field over the 2DEG device.
\\\

%\section{Conclusion}
In conclusion, we have shown that efficient field effect control of the carrier density in a superconducting 2DEG formed at the AlO$_x$/KTaO$_3$~(111) interface may be achieved using accessible fabrication methods: with a stainless steel shadow mask using no chemical surface treatment, and making use of classic UV lithography techniques in combination with ion beam milling at low temperature or deposition of an insulating hard mask. These fabrication techniques may help design more complex devices to control and harness the properties of the superconducting phase of the interface.

%\medskip
%\textbf{Supporting Information} \par %Please delete the Suppporting Information statement if it is not applicable. Please supply Supporting Information in another file. Supporting information should not be provided in .tex format
%Supporting Information is available from the Wiley Online Library or from the author.

% Acknowledgements
\medskip
\textbf{Acknowledgements} \par %delete if not applicable))
This work was supported by the ANR QUANTOP Project-ANR-19-CE470006 grant, by the QuantERA ERA-NET Cofund in Quantum Technologies (Grant Agreement N. 731473) implemented within the European Union’s Horizon 2020 Program (QUANTOX). We thank Marco Salluzzo for useful comments.

% References
\medskip
% Use the following code if you wish to generate your bibliography with BibTeX;
% replace the string "MSP-template" below with the name(s) of
% the BibTeX data base(s) you want to use.
% The resulting bibliography-output (the content of the .bbl file)
% must be pasted back into this file before submission.
% Please also include your BibTeX data base file(s) in your submission
% so that we can re-run BibTeX if necessary.

%\bibliographystyle{MSP}
%\bibliography{2DEG_patterning_bib}

\end{document}